\documentclass[sigconf,natbib=false]{acmart}
\usepackage{array}
\usepackage{subcaption}
\usepackage{siunitx}
\usepackage{booktabs}
\usepackage{adjustbox}
\usepackage[utf8]{inputenc}
\AtBeginDocument{%
  }

\setcopyright{acmlicensed}

\copyrightyear{2026}
\acmYear{2026}
\setcopyright{cc}
\setcctype{by-nc-nd}
\acmConference[SAC '26]{The 41st ACM/SIGAPP Symposium on Applied Computing}{March 23--27, 2026}{Thessaloniki, Greece}
\acmBooktitle{The 41st ACM/SIGAPP Symposium on Applied Computing (SAC '26), March 23--27, 2026, Thessaloniki, Greece}
\acmPrice{}
\acmDOI{10.1145/3748522.3779988}
\acmISBN{979-8-4007-2294-3/2026/03}

\RequirePackage[
  datamodel=acmdatamodel,
  style=acmnumeric,
  ]{biblatex}

\usepackage{array}
\begin{document}

\title{Examining Software Developers' Needs for Privacy Enforcing Techniques: A survey\\}
\renewcommand{\shorttitle}{Developers' Needs for Privacy Enforcing Techniques}

\author{Ioanna Theophilou}
\affiliation{%
  \institution{University of Cyprus}
  \city{Aglantzia}
  \country{Cyprus}}
\email{theophilou.ioanna@ucy.ac.cy}

\author{Georgia M. Kapitsaki}
\affiliation{%
  \institution{University of Cyprus}
  \city{Aglantzia}
  \country{Cyprus}}
  \email{gkapi@ucy.ac.cy}

\renewcommand{\shortauthors}{I. Theophilou and G. M. Kapitsaki}

\begin{abstract}
Data privacy legislation, such as GDPR and CCPA/CPRA, has rendered data privacy law compliance a requirement of all software systems. Developers need to implement various kinds of functionalities to cover law needs, including user rights and law principles. As data compliance is tightly coupled with legal knowledge, it is not always easy to perform such integrations in software systems. Prior studies have focused on developers' understanding of privacy principles, such as Privacy by Design, and have examined privacy techniques used in the software industry. Nevertheless, emerging developer needs that can assist in privacy law compliance have not been examined but are useful in understanding what development automation tools, such as Generative AI, need to cover to make the compliance process more straightforward and seamless within the development process. In this work, we present a survey that examines the above needs with the participation of 68 developers, while we have examined which factors affect practitioners' needs. Most developers express a need for more automated tools, while privacy experience increases practitioners' concerns for privacy tools. Our results can assist practitioners in better positioning their development activities within privacy law compliance and point to an urgent need for privacy facilitators. 
\end{abstract}

\ccsdesc[300]{Security and privacy~Software and application security}
\ccsdesc[500]{Software and its engineering~Software design engineering}
\ccsdesc[300]{Software and its engineering~Software development techniques}
\ccsdesc[500]{Security and privacy~Privacy protections}
\keywords{Data Privacy Law, GDPR, Developers' Survey, Software Law Compliance.}

\maketitle

\section{Introduction}

The General Data Protection Regulation (GDPR) put into practice since 2018 includes a set of rules on the collection, storage and processing of personal data~\cite{eugdpr}. The regulation brought significant changes to both research and industry, driving significant changes across all stages of the Software Development Life Cycle (SDLC)~\cite{Leite2021}. The concept of `Privacy by Design' that is mentioned in GDPR and applied by practitioners was first introduced in 1990 as a methodology that aims to incorporate the concept of privacy from the initial stages of designing a software system~\cite{trujillo19}. Beyond laws and policies, researchers and engineers have tried to make Privacy by Design more practical with privacy patterns offering reusable solutions to common design problems~\cite{Colesky2016}. More recently, Pallas et al. suggested a road-map that brings privacy principles closer to real industry practice, showing continued efforts to bridge the gap between theory and software development~\cite{pallas2024privacy}.  

Despite the above, the increasing regulatory pressures and the critical role of privacy in software development, developers are often expected to make privacy decisions without formal training or expertise, including also no training on legal practices. Through the current work, we introduce a survey that examines developer privacy perceptions and barriers to building privacy-aware and GDPR-compliant systems, while investigating how developers enforce privacy practices across the Software Development Life Cycle. The aim is to understand current needs and investigate the feasibility and attractiveness of alternative practical solutions. Existing survey-based studies on privacy with software developers have only examined one dimension of their interaction with privacy, for instance: behaviors, current practices, or needs~\cite{franke2024exploratory,prybylo2024evaluating,senarath2018PrExpect}, without integrating these perspectives, while they do not investigate developers' practical needs. In the current study, we examine how the above dimensions are interconnected.

Through the survey, we aim in answering the following main research problem: \textit{What are software developers’ needs for law-compliant privacy-enforcing techniques?} To achieve this, we define the following four \textit{Research Questions (RQs)}: \textbf{(1) RQ1}. \textit{Which needs do software developers have for creating privacy-aware code?} \textbf{(2) RQ2}. \textit{How do developers' activities and experience relate to their needs for privacy tools?} \textbf{(3) RQ3}.  \textit{How do demographics and organization size affect the needs for privacy tools?} \textbf{(4) RQ4}. \textit{How would software developers react in privacy-leak realistic scenarios?} For the first three RQs, we relied both on general and technical needs, while the fourth RQ was added in order to gain a better insight on practitioners current understanding and find basic solutions applied. We analyzed 68 valid developers' responses and found as main outcomes that there is a need for more automated tools, while privacy experience increases practitioners concerns for privacy tools.

The remainder of the paper is structured as follows. Section 2 presents related work in the area, while the survey creation process is detailed in Section 3. Quality assurance and analysis is introduced in Section 4 and main results are presented and discussed in Section 5. Section 6 describes briefly implications and threats to validity, and finally Section 7 concludes the work.

\section{Related Work}

\textbf{Privacy tools for software development}. The need for automated detection of potential privacy violations in software code before it reaches production has been highlighted in~\cite{mao2024privacycat}. The authors developed PrivacyCAT, a code analysis tool designed to identify privacy vulnerabilities and alert developers during the early stages of software development. The study concludes by emphasizing the value of automated privacy vulnerability detection for ensuring privacy compliance and reducing risk. Tahaei et al. analyzed developers' approaches to privacy tasks in software development, finding a focus on regulatory compliance and confidentiality in 119 Stack Overflow answers~\cite{Tahaei20}. The results suggest promoting overlooked strategies through tools and enhancing support for managing third-party data practices. Tool-based formal approaches have also been proposed for integrating GDPR principles into the SDLC~\cite{vanezi2020dialogop}.

\textbf{Developer surveys}. What prevents software developers from incorporating GDPR principles into their everyday development practices has been investigated in~\cite{alhazmi2021gdpr}. The authors conducted a survey with 22 developers who were asked to study a scenario and its accompanying UML diagrams before participating in interviews where they explained how they would implement specific GDPR laws within the scenario. Key findings include that developers lack clear techniques for implementing principles like Storage Limitation, Purpose Limitation, and Accuracy, and they are often confused by Lawfulness of Purpose, Fairness, and Transparency. Many developers are unfamiliar with the GDPR legislation, and some do not consider privacy their responsibility. The study also found a lack of resources, tools, and implementation guidelines, and that privacy expectations often vary depending on organizational budgets.

In the literature, there are five prior works that align more with our survey. Prybylo et al. conducted a study as a way to understand software developers' experiences with GDPR compliance challenges~\cite{prybylo2024evaluating}. Through this survey with 56 participants, the authors identified gaps in privacy practices among software teams and conclude that role-dependent solutions should reply to all privacy challenges faced. Senarath et al. conducted a focus group-type survey with 151 participants, 30 of whom were software developers~\cite{senarath2019Model}. The paper studied how people feel about their privacy when software collects different kinds of personal data. In the focus group, participants had to review 30 scenarios in order to create 30 privacy risks. The authors proposed a model aimed for software developers to take such metrics.

In another work by the same authors, software developers were put in different scenarios and roles in order to define users' expectations around privacy \cite{senarath2018PrExpect}. This survey included 56 participants in total, 36 of whom were software developers. The authors also tested developers' perception with basic questions on privacy and the GDPR. The study revealed that both entities' (developers and end-users) expectations on privacy matters matched in the case where developers were answering from a `user point-of view.'
 
Franke et al. asked Software Developers on currently used techniques for GDPR compliance while trying to understand how laws affected Open-Source Software (OSS) development~\cite{franke2024exploratory}. They conducted a survey with 56 participants where they also included some interview questions to gather insights on software developers' experiences. Their results revealed that the regulation made OSS development difficult, time-consuming and confusing for software developers. Another work tried to understand what Developers perceive as Privacy Threats in Social Coding Platforms (SCPs)~\cite{ferreyra2023SCPs}. With 105 developers as participants, the authors tested their perception using some `privacy attack scenarios.' Results showed that developers do not actually feel threatened of being identified rather that hidden inferences being revealed. 

In order to provide a summary and compare the current work with prior surveys, we identified five categories of context types these works refer to. The categories are: (1) Awareness, (2) Perception/Comprehension, (3) Developer's Behavior, (4) Implementation Challenges for GDPR compliance, and (5) Current Practices. Table~\ref{tab:relatedworkCompare} shows that existing literature focuses on some areas in understanding software developers' point of view in terms of privacy tasks. In our work, we cover all categories in the survey questions but due to limited space, for the analysis we focus mainly on developers' specific general and technical needs for privacy enforcement that is an area missing in all prior works. Only~\cite{prybylo2024evaluating} discusses this as a take-away message. We plan on covering survey results from the other categories on future publications. 

\begin{table}[h!]
\centering
\begin{tabular}{|l|c|c|c|c|c|c|c|}
\hline
\textbf{Category} & \cite{alhazmi2021gdpr} & \cite{prybylo2024evaluating} & \cite{senarath2019Model} & \cite{senarath2018PrExpect} & \cite{franke2024exploratory} & \cite{ferreyra2023SCPs} & Our\\ \hline
Awareness & &\checkmark  & &  &  &\checkmark  &\checkmark\\ \hline
Perception & &  &\checkmark  & \checkmark &\checkmark  &\checkmark  &\checkmark\\ \hline
Developer’s  & &  &  & \checkmark &  &  &\checkmark\\ 
Behavior & & &  &  & &  &\\ \hline
GDPR Implement. & \checkmark & \checkmark &  &  & \checkmark  &  &\checkmark\\ 
Challenges & & &  &  & &  &\\ \hline
Current Practices & & \checkmark &  &\checkmark  &\checkmark  &  &\checkmark\\ \hline
Needs & &part. &  &  &  &  & \checkmark\\ \hline
\end{tabular}
\caption{Areas targeted in related work surveys.}
\label{tab:relatedworkCompare}
\end{table}
\section{Survey Creation Process and Structure}
\subsection{Main survey design process}

In order to design the survey, we reviewed the respective prior works (Table~\ref{tab:relatedworkCompare}) and their survey questions and identified gaps that need to be addressed via a new survey (mainly developer needs). One main goal was to address all categories instead of individual categories (awareness, perception,behavior,GDPR challenges, current practices) targeted in the existing literature in order to get a more holistic view, while focusing on specific needs that are now missing. Through our survey, participants were required to think about how they currently enforce privacy and reflect on their current methods, express their challenges and be more aware of what could be done to make their day-to-day privacy tasks easier while publishing products that comply to privacy legislation. We followed the design guidelines of Altschuld et al. who mention that surveys whose main goal is the ‘Needs assessment’ always have to answer two critical questions: The (1)’ What is’ (=what currently exists) and (2) ‘What should be’ (=what is actually needed)~\cite{Altschuld22}. We included thus, a section asking for ‘current practices’ (e.g. if there is a Privacy Officer position in the organization) and another section on ‘general needs and technical needs.’ Concerning current understanding, we also included at the end of the survey four realistic scenarios that reveal privacy vulnerabilities. We asked developers what they would do in each case, and what they think would be the best option as a solution. The final survey sections are presented later in this section.

In the questions on needs, we asked developers if they agree on 20 statements (10 for general and another 10 for technical needs) that were derived as described next. On the general needs questions, we focused on addressing all main gaps found in the literature. Many studies reveal that Privacy by Design remains supported only in theory due to lack of tools~\cite{Hadar18,Tahaei23}. We included this by specifically asking developers if they believe that there is a need for developer-friendly tools that support Privacy by Design. Prior work also showed that privacy frameworks play a huge role in compliance and adoption~\cite{Deng11}. We asked developers if they believe frameworks would be beneficial for their SDLC. Software Engineering research has also found that developers have difficulties translating privacy requirements into engineering techniques~\cite{Senarath18Investigation}. The adoption of Privacy-Enhancing Technologies (PETs) is not common practice yet due to usability and integration barriers \cite{Hasani23}. To cover this, we added relevant statements asking for participants views in questions, e.g. ”\emph{I need clearer requirements related to user privacy from stakeholders}” and “\emph{I would use Privacy-Enhancing Technologies if they were easier to integrate}”. 

The same strategy was followed to create the more technical questions. It has been discussed that developer-friendly APIs do not exist to accommodate common privacy tasks like consent tracking~\cite{Tahaei2022understanding}. Regarding automation and use of reusable components on privacy tasks, many studies have proven the lack of such tools~\cite{GUBER2024107351,Kühtreiber22,Sangaroonsilp25}. To cover this, we added statements like: “\emph{There is a lack of modular, reusable components for implementing common privacy tasks (e.g., anonymization, access control)},” “\emph{Tools that help ensure compliance with data-minimization and purpose-limitation principles would be useful}” and “\emph{Automated tools that detect privacy risks in code would significantly help our workflow}.” Before finalizing the survey structure, we conducted a pilot survey as detailed next.

\subsection{Pilot survey}

The purpose of the pilot survey was to get feedback from experts before disseminating the final survey to development teams. We asked all pilot participants to give us feedback on the questions and answer the survey themselves. We aimed at addressing any gaps in terms of comprehension and relevance in the questionnaire from the recruited participants. We were able to extract useful information about how targeted the questions are from the way participants answer. For the pilot study, we recruited participants with different backgrounds, ages and expertise to ensure that the results would be unbiased. The general research was presented as an early-stage idea in an event organized by the Department of Computer Science of University of Cyprus.
In this event, industrial affiliates of the Department were present (CEOs, CTOs, software developer team leads, etc.). The pilot survey was presented to a small pool of attendees within that group along with some researchers. 
Participants showed interest in the research and requested to be given the final format of the survey to either respond to or disseminate within their development teams. 
 
The feedback we received verified that questions were fairly clear and understandable. Feedback included also some grammar corrections, typos and other notes in terms of relevance and suggestions on some questions. It was decided to remove some questions from the initial design, since they were found to be non-beneficial for our purposes.For example, the pilot survey included the following question: `3.4 Please select all personal data from the list below that are considered sensitive.' The question addresses solely awareness which did not fit the purpose of the survey. The question was found to be irrelevant and it was then removed.  We also enhanced the survey with explanations of some terminologies that not all pilot participants were familiar with. Suggestions from participants included changes in the scenario phrasing, as well as contextual enhancements of the scenarios. For example, one participant mentioned that it would be beneficial if we included some detailed information of Software Development Life Cycle in our scenarios. We accommodated that by including a scenario that concerns a specific Sprint cycle. We did that to make the respondent understand that in that specific scenario the team is in the early stages of the design phase which might play a role on their response. From the analysis of the preliminary responses and the feedback gathered, we extended the survey with more privacy-relevant development scenarios, using open-ended questions. Demographics and questions regarding current practices that were part of the original survey remained the same.

\subsection{Informed consent form}

The final version of the study has received approval by the National Bioethics Committee of Cyprus.\footnote{\url{https://www.bioethics.gov.cy}} The study includes a thorough informed consent form. Through the consent form, participants were informed that data collected are anonymous and all information will be safeguarded subject to any legal requirements. They were also informed that the participation is voluntary, and that they can withdraw from it at any time without having to explain why. We ensured that we comply to the GDPR rights by following practices for the \emph{Right to Access}, \emph{Purpose Limitation}, \emph{Data minimization}, \emph{Rectification} and \emph{Data Erasure}. The complete survey structure is detailed next and is available online: \url{https://doi.org/10.6084/m9.figshare.30374404}.

\subsection{Final survey structure}

The survey is divided into six main sections (including the consent form section). Following as mentioned previously the design guidelines of Altschuld et al.~\cite{Altschuld22}, questions included whether participants have a standardized procedure for privacy compliance, what tools they use for this purpose (if any), how confident they are with their current practices and what challenges they had to face in a privacy-crisis. We also examined their familiarity with privacy laws, user rights and whether law-compliant code is part of their everyday tasks. Participants were then presented with the general technical needs. Lastly participants were given four realistic scenarios to act upon. The specific sections of the final survey are the following:

\textbf{Demographics}. The following demographic data were collected in the second section of the survey: (1) age group, (2) gender, (3) country of residence, (4) highest degree of education, (5) role in the development team, (6) years of development experience, (7) company size of current position.

\textbf{Current Practices}. In the current practices section, we collected information regarding daily tasks and procedures followed in the software development team: (1) if the company has a Chief Privacy Officer, (2) if developers are personally expected to create privacy policies, (3) if the participants have ever created a privacy policy (if yes they could state the policy generation methods and tools used along with the challenges that come with it), (4) which SDLC they use, (5) how confident they are with their current privacy practices, (6) if they use any Privacy Enhancing Technologies, and (7) if they use forums or any AI tools to solve privacy-related matters in their everyday life (and if yes, which ones).

\textbf{Awareness/Familiarity}. In this section, we assessed familiarity with privacy legislation, user rights and Privacy Impact Assessments (PIA). We also asked participants if privacy-compliant code is part of their daily tasks. 

\textbf{Addressing the privacy needs}. We created a pool of statements grouped as `general needs' and `technical needs' where we ask to which level the participants agree on each statement (using a 5-Likert scale). We have also included three realistic scenarios that promote critical thinking on how to solve privacy-related problems that may arise:

\emph{Scenario 1} presented a REST API commit that reveals risky information in the response. We asked developers how they would modify the response, whether they would prefer an automated procedure that modifies the response for them and what their ideal automated procedure would be. \emph{Scenario 2} described the functionality requirement of automation of administration account reviews. These are the periodic examination and verification procedures of permissions and rights granted to the system administrators. The scenario mentioned an SQL SELECT statement that includes some Personally Identifiable Information. Developers were asked in which part of their code 
they would require users' consent and how often. \emph{Scenario 3} described a situation where privacy risks appear after deployment. It mentioned that the developer has integrated a third-party Software Development Kit (SDK) to track user engagement in a mobile app. The developer deployed the SDK and then realized that it collects location data by default. We asked developers whether they would use a tool that detects such issues and how.

\textbf{Detection of Privacy Leak}. In this section, we wanted to introduce the notion of Privacy as Code~\cite{ferreyra2025good}. Coding with privacy in mind would prevent many privacy leaks. We presented participants with three realistic questions all related to a specific case study. The case study showed a pseudo-code implementation and put participants in the position of thinking solutions in how to make this implementation more private. Details on the SDLC were also presented to the user to make this more realistic.

The case study (\textit{Scenario 4}) included two questions around the following situation: Developers (participants) are working at a company building SafeHealth, a web-based eHealth platform. Patients (which are the direct clients) use the platform to fill out health questionnaires, book appointment and consult doctors through video calls. We mentioned that the SDLC used is Scrum with 2-week sprints. We also mentioned that their task was to implement secure storage of user-submitted health questionnaires and we presented participants with a pseudo-code initial implementation in Sprint 3 (early phase in the development). Participants were asked about privacy risks (multiple choice question) and how they would incorporate privacy controls into the Agile workflow.


\section{Quality Assurance and Data Analysis Methods}

A methodology was followed to ensure quality before and after the dissemination of the survey based on established methods from the literature described briefly next. In~\cite{Iwaya_2023}, in order to ensure quality the authors focus on diverse participant sampling to reduce bias. They also used thematic analysis with systematic coding and validation. In~\cite{ferreyra2023SCPs}, it is made sure that the questionnaire was built from established constructs in the literature for validity. The authors also used hypothetical attack scenarios to test consistency of participant responses. Prybylo et al., on the other hand, started with a pilot test, did pre-screening through Prolific website and after the results were in, they evaluated free-text manually to filter out the non-relevant target group~\cite{prybylo2024evaluating}. They also filtered out incomplete surveys.
 

Our methodology 
is divided in two sections: the Pre-testing and Validation section (= before the dissemination) and the Data Cleaning section (=procedure done after dissemination). Through the Validation and Pre-testing procedure, we firstly researched similar studies (Table~\ref{tab:relatedworkCompare}) and established a list of validated questions from the read literature. We then conducted a pilot study (also described in the previous sections) where we got immediate feedback and re-designed parts of the survey structure and content. We then made sure to make efforts to recruit a diverse group of participants by disseminating the survey through eight public groups on LinkedIn and six groups on Reddit along with manual sharing via email communication (to company contacts and to the industrial affiliates of our Department). Data Cleaning was firstly done through the translation of all answers to English, followed by the manual exclusion of participants that did not fit in our target group. We then extracted and excluded spam answers through manual walkthrough of the free-answer questions. The above were performed on the 72 responses we received between August and October 2025. Of those, three were identified as spam and one stated `no' in consent, so the final number of analyzed responses was 68. There were no cases of replies with participants not fitting in our target group, while there were two replies in Chinese that were translated to English.

The mathematical survey design with statistical tests that we adopted includes examining correlations and associations, along with group differences. For the statistical analysis, we used IBM SPSS where we run descriptive and non-parametric k-independent samples tests whereas scenarios analysis was performed in Python. 

\section{Survey Results and Discussion}

\subsection{General demographics}

\begin{figure}
    \centering
    \includegraphics[width=0.80\linewidth]{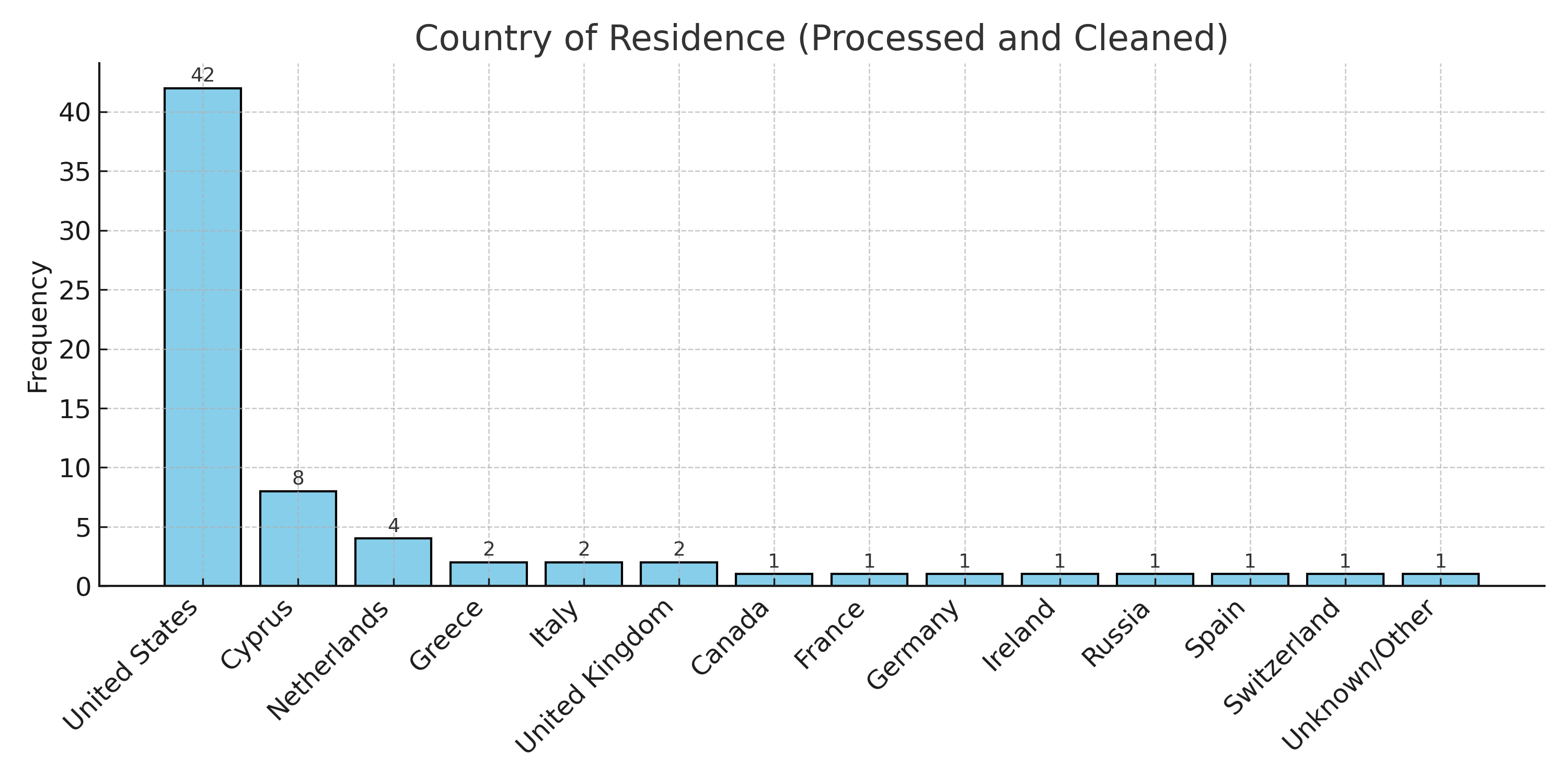}
    \caption{Survey participants country of residence.}
    \label{fig:country}
\end{figure}

Among the 68 participants, 33 (47.9\%) participants identified as female and 35 (52.1\%) as male. Although the given options in the survey form were five different age groups, we got responses from the following groups: "18-28" (28.2\%), "29-39" (53.5\%) and "40-50" (18.3\%). Figure ~\ref{fig:country} summarizes participants' countries. The majority of respondents are from the United States (42), followed by Cyprus (8), the Netherlands (4), the United Kingdom (2), Greece (2), and Italy (2). Countries with only one response are:  Canada, France, Germany, Ireland, Russia, Spain, and Switzerland. Additionally, one respondent identified their region as the Middle East (instead of providing a specific country name).

\subsection{RQ1. Which needs do software developers have for creating privacy-aware code?}

Concerning technical needs, developers indicated the `\emph{lack of modular, reusable components for implementing common privacy tasks}' (52/68 participants agree or agree strongly that is is needed) and the `\emph{automated tools to detect privacy risks}' (50/68 participants agree or agree strongly) as the top privacy needs. On general needs, `\emph{reusable privacy design patterns}' (57/68 participants agree or agree strongly that is is needed) and `\emph{stronger collaboration between legal experts and developers}' (51/68 participants agree or agree strongly) were found especially useful. The descriptive statistics for all general and technical needs along with respective Item IDs used for the needs for brevity purposes in the remainder of the text are shown in Table~\ref{tab:item_summary}.
\renewcommand{\arraystretch}{0.80}

\begin{table*}[htbp]
\centering
\small
\setlength{\tabcolsep}{4pt}
\caption{Descriptive statistics of top privacy needs of participants.}
\begin{tabular}{lp{14.1cm}rr}
\toprule
\textbf{Item ID} & \textbf{Question} & \textbf{Mean} & \textbf{Std. Dev.}\\
\midrule
\multicolumn{4}{@{}c}{General needs} \\
\midrule
4.1.1 & There is a need for more developer-friendly tools that support privacy-by-design. &4.14 &0.731 \\
4.1.2 & There is a lot of documentation/practical guides on adding privacy principles in code.&3.84 &1.031 \\
4.1.3 & My team would benefit from privacy frameworks integrated into our development process.&4.06 &0.790 \\
4.1.4 & I need clearer requirements related to user privacy from stakeholders.& 4.19&0.868 \\
4.1.5 & There are many training opportunities on privacy in software engineering. &3.71 &0.947\\
4.1.6 & I would use privacy-enhancing technologies more often if they were easier to integrate into existing systems.&4.21 &0.873 \\
4.1.7 & My/Our development process lacks automated tools for detecting potential privacy violations.&3.94 &0.912 \\
4.1.8 & There is a need for stronger collaboration between legal experts and developers for privacy.&4.01 &1.000 \\
4.1.9 & I feel the need for reusable privacy design patterns.&4.16 &0.803 \\
4.1.10 &Organizational support (e.g., management prioritization, dedicated roles) is needed to better address privacy in our software.& 4.03&0.810 \\
\midrule
\multicolumn{4}{@{}c}{Technical needs} \\
\midrule
4.2.1 & There is a need for APIs/libraries that simplify the implementation of privacy-enhancing features.&4.12 &0.802 \\
4.2.2 & I would use more privacy tools if they were better documented and easier to integrate.&4.15 & 0.868\\
4.2.3 & Our current development tools (IDEs, CI/CD, testing frameworks) have sufficient built-in support for privacy checks. &3.76 &1.009\\
4.2.4 & Automated tools that detect privacy risks in code would significantly help our workflow. &4.06 &0.790 \\
4.2.5 & We need privacy enforcement mechanisms embedded in the software architecture from the start.&4.21 &0.907 \\
4.2.6 & It is difficult to implement privacy features without compromising system performance.& 3.85&0.902 \\
4.2.7 & There is a lack of modular, reusable components for implementing common privacy tasks (e.g., anonymization, access control). &4.09 &0.958\\
4.2.8 & Tools that help ensure compliance with data minimization and purpose limitation principles would be useful&3.88 &0.856\\
4.2.9 & There is sufficient guidance on how to apply PETs (e.g., encryption, differential privacy) within software systems. &3.79 &1.127\\
4.2.10 & Privacy-related logging, auditing, and data access tracing should be easier to implement at the code level&4.01 &0.954 \\
\bottomrule
\end{tabular}
\label{tab:item_summary}
\end{table*}

\subsection{RQ2. How do developers' activities and experience relate to
their needs for privacy tools?}

In this RQ, we examined the participants view on privacy tool needs in relevance to their activities and experience running for this purpose a number of statistical tests between the needs of Table~\ref{tab:item_summary} and earlier parts of the questionnaire. A Cochran-Armitage trend test (SPSS Linear-by-Linear Association) showed a statistically significant positive trend between existing use of PETs by the participants and agreement that there is a lack of automated tools ($\chi^2(1)=6.744$, $p-value=0.009$). This means that people with experience with PETs are increasingly likely to agree with the statement. In the same direction, a relation was found between confidence with privacy measures and the perceived need for built-in privacy support tools. Somers'd indicated a moderate positive ordinal association ($D=0.27$, $SE=0.095$, $p-value=0.005$).

In order to examine associations and correlations regarding different participants' roles and privacy needs, we ran the Row-columns-effect model. As shown in Table \ref{tab:sig_only_numeric}, the role with the most significant p-value is the \emph{Information Security/Privacy area}-related role
. Results that were not statistically significant are not reported in the table. We also expected to find correlations between awareness of data protection rights and need for data minimization tools. This was unexpectedly not associated though ($p-value=0.08$, Somers' d of $0.37$). So the null hypothesis stands in this case, with all groups having similar agreement in terms of need. The Shapiro–Wilk test was used to assess normality and results indicated that all variables significantly deviated from a normal distribution ($p-value<0.001$). So forth we are using next non-parametric tests when examining group differences (Kruskal Wallis and Mann-Whitney U tests). 

\begin{table}[htbp]
\centering
\small
\setlength{\tabcolsep}{6pt}
\caption{Row-columns-effect model results with statistical significance for roles and needs.}

\begin{tabular}{p{2cm}cp{1.5cm}rr}
\toprule
\textbf{Role} & \textbf{PairID} & \textbf{Test} & {\textbf{$\chi^2$ }} & {\textbf{p-value}} \\
\midrule
Team Leader & 1  & Pearson & 8.97 & 0.030 \\
Software Tester & 1  & Linear-by-Linear trend & 4.94 & 0.026 \\
QA Engineer & 2 & Linear-by-Linear trend & 5.13 & 0.023 \\
Information Security/Privacy& 2 & Pearson & 9.64 & 0.022 \\
Information Security/Privacy& 2 & Linear-by-Linear trend & 8.49 & 0.004 \\
Information Security/Privacy& 3 & Linear-by-Linear trend & 4.46 & 0.035 \\
\bottomrule
\end{tabular}

\vspace{1pt}
\raggedright
\footnotesize

\textbf{PairID}: 1=Need for privacy-by-design tools,\; 2=Benefit from privacy frameworks in SDLC,\; 3=Need for clearer privacy requirements.\\
\label{tab:sig_only_numeric}
\end{table}

The role in the survey was a nine-option multiple choice question. Since values could be combined to identify someone's full role in the team, in SPSS we implemented this by having each role as a different variable with the values 0 as Not Selected and 1 as Selected. We run a Mann-Whitney U non-parametric test for all roles as factors and all privacy needs items (general and technical) of Table~\ref{tab:item_summary} as dependent variables. Administrator and Software Design/Architecture roles are not shown, since they revealed no significant results.

\begin{table}[htbp]
\centering
\small
\setlength{\tabcolsep}{3pt}
\caption{Significant Mann--Whitney U test results with mean ranks for Selected and Not selected groups.}
\begin{tabular}{lcrrr}
\toprule
\textbf{Role} & \textbf{ItemID} & \textbf{p-value} & \textbf{Selected} & \textbf{Not} \\
&&&&\textbf{selected} \\
\midrule
Team Leader & 4.2.2 & 0.003 & 22.38 & 38.23 \\
Team Leader & 4.1.7 & 0.024 & 25.31 & 37.33 \\
Scrum Master & 4.2.1 & 0.005 & 14.33 & 36.45 \\
Scrum Master & 4.2.10 & 0.016 & 17.00 & 36.19 \\
Scrum Master & 4.2.5 & 0.022 & 18.17 & 36.08 \\
Product Manager & 4.1.2 & $<0.001$ & 45.42 & 28.55 \\
Product Manager & 4.1.8 & 0.002 & 44.17 & 29.23 \\
Product Manager & 4.1.5 & 0.015 & 41.96 & 28.31 \\
Product Manager & 4.1.10 & 0.024 & 41.42 & 30.73 \\
Product Manager & 4.2.6 & 0.041 & 40.69 & 31.13 \\
QA Engineer & 4.2.7 & 0.004 & 46.80 & 31.02 \\
QA Engineer & 4.1.3 & 0.025 & 43.87 & 31.85 \\
Software Developer & 4.2.2 & $<0.001$ & 44.23 & 27.69 \\
Information Security/Privacy & 4.1.2 & $<0.001$ & 48.76 & 28.13 \\
Information Security/Privacy & 4.2.3 & $<0.001$ & 46.69 & 29.05 \\
Information Security/Privacy & 4.1.3 & 0.004 & 44.12 & 30.20 \\
Information Security/Privacy & 4.2.6 & 0.003 & 44.57 & 30.00 \\
Information Security/Privacy & 4.2.9 & 0.003 & 44.69 & 29.95 \\
Information Security/Privacy & 4.1.8 & 0.027 & 42.00 & 31.15 \\
Information Security/Privacy & 4.1.9 & 0.034 & 41.48 & 31.38 \\
Information Security/Privacy & 4.1.10 & 0.042 & 41.36 & 31.44 \\
Software Tester & 4.1.1 & 0.021 & 49.50 & 32.78 \\
\bottomrule
\end{tabular}

\vspace{2pt}
\raggedright
\label{tab:role_numeric_sorted_byrole}
\end{table}

Main results for the above tests from Table~\ref{tab:role_numeric_sorted_byrole} are the following: (1) Non-team leads respondents had higher mean ranks for lack of automated tools, indicating stronger agreement with this statement. (2) There is a statistically significant difference between Team leaders on the statement “\emph{I would use more privacy tools if they were better documented and easier to integrate}” ($p-value=0.003$ for Mann-Whitney U test). (3) It is more likely for non-Scrum Masters to agree positively in needing APIs for simpler implementation of privacy features, in needing privacy enforcing mechanisms and that logging and auditing should be easier to implement at code-level. (4) Product Managers seem more aware of documentation guidelines for privacy principles in code ($p-value<0.001$) and (5) on training opportunities ($p-value=0.015$). (6)  Product Managers seem to also agree more ($p-value=0.002$) on the need for stronger collaboration between legal experts and developers along with the need for organizational support ($p-value=0.0024$) to address privacy in software. (7) Product Managers seem more concerned on privacy over system performance ($p-value=0.041$). (8) QA Engineers seem to agree more on the benefit of integrated privacy frameworks ($p-value=0.025$) and the lack of reusable components on privacy tasks ($p-value=0.004$). (9) Software Developers have generally stated more than others that they are very willing to use more privacy tools knowing that they are easy to integrate ($p-value=<0.001$) showing a clear need. (10) Information Security/Privacy Developers express a strong need for integrating privacy principles in code ($p-value=<0.001$) and a strong (11) perceived benefit of framework integration ($p-value=0.004$). (12) They identify the lack of automated tools for privacy leak detection ($p-value=0.010$) and they promote (13) collaboration with legal experts ($p-value=0.027$) and (14) organizational support in general ($p-value=0.042$). Finally, they also show more interest than other roles in the development life cycle for (15) reusable privacy patterns ($p-value=0.034$). 

\subsection{RQ3. How do demographics and organization size affect the
needs for privacy tools?}

Concerning the effect of demographics on the participants needs, main results using numeric variables are shown in Table~\ref{tab:kruskal_numeric} and include the following (Note: groups (g1–g6) correspond to each factor’s original categories e.g. Age-g1=`18-28'): 
(1) Post-hoc testing revealed that participants from smaller companies were less likely to feel there is not enough guidance available on PETs. (2) Overall, participants tend to disagree or remain neutral about the idea that there are many training opportunities on privacy, regardless of experience. (3) Different experience levels revealed different levels of agreement on lacking of automated tools for privacy leak detection. Mid-level professionals (6–10 years of experience) showed the highest agreement with this statement. (4) More experienced participants seem more confident on the current guidance around PETs. (5) Mid-level participants tend to agree more that automated tools would help their workflow  according to median values observed. (6) Mid-level participants tend to recognize more the need for a stronger collaboration with legal experts.

\begin{table}[htbp]
\centering
\small
\setlength{\tabcolsep}{3pt}
\caption{Significant Kruskal--Wallis results with group frequencies ($>$Median/$\le$Median).}
\begin{tabular}{p{1.2cm}crp{4.5cm}}
\toprule
\textbf{Factor} & \textbf{ItemID} & \textbf{p-value} & \textbf{Groups ($>$M/$\le$M)} \\
\midrule
Age & 4.1.5 & 0.044 & g1: 4/15; g2: 10/26; g3: 0/13\\
Age & 4.2.7 & 0.026 & g1: 3/16; g2: 17/19; g3: 7/6;\\
Company size & 4.2.9 & 0.039 & g1: 0/2; g2: 1/5; g3: 1/4; g4: 1/9; g5: 6/8; g6: 12/19 \\
Experience & 4.1.5 & 0.033 & g1: 1/10; g2: 2/16; g3: 6/18; g4: 5/10 \\
Experience & 4.1.7 & 0.021 & g1: 2/9; g2: 4/14; g3: 11/13; g4: 3/12 \\
Experience & 4.1.8 & 0.046 & g1: 2/9; g2: 3/15; g3: 14/10; g4: 6/9 \\
Experience & 4.2.3 & 0.027 & g1: 1/10; g2: 4/14; g3: 10/14; g4: 2/13 \\
Experience & 4.2.9 & 0.013 & g1: 2/9; g2: 3/15; g3: 12/12; g4: 4/11 \\
\bottomrule
\end{tabular}
\vspace{2pt}
\raggedright
\label{tab:kruskal_medians_compact}
\label{tab:kruskal_numeric}
\end{table}

\subsection{RQ4. How would software developers react in privacy-leak realistic scenarios?}

In order to analyze participants' responses on the vulnerability scenarios, we used an automated approach, where through n-gram analysis \footnote{N-gram analysis is a technique that breaks text into sequences of N items. Its goal is to find any patterns.}, we extracted the top 2-grams, 3-grams and 4-grams. For pre-processing purposes, we used a list of stop-words in English (e.g. a, about, above, after, again, against), removing generally used terms from our corpus.

\textbf{Scenario 1} (REST API commit that reveals risky information). The most popular response by participants (in 3-gram phrases) was `remove sensitive fields' (appearing 12 times). Two phrases were found to be the second most popular responses (each appearing 6 times): `filter sensitive fields',  `integrate CI / CD' (=Continuous Integration/Continuous Delivery). Other popular solutions proposed by developers were: `statistic code analysis', 'name automation' and `schema validator tools'. From the 4-gram phrases we extracted popular responses like: `immediately keep necessary info' as solutions and `like spectral scan OpenAI' as currently used tools. From the responses, we can see that developers prefer \textbf{proactive} integration. They emphasize on \textbf{data minimization} (`remove' `filter' `keep only') and that the interest in automated solutions like schema validators or code scanners is high. 

\textbf{Scenario 2} (Procedures of permissions granted to the system administrators). Results processing shows that the most popular answers are the following: ` CSV offline processing', `time data stored' `data stored processed' and `data stored shared'. The most repeated word was `processing' (37 mentions), `storage' (24 mentions) and `sharing' (20 mentions). From the 4-gram results, we can see that most frequent appearing phrases in participants replies are: `time data stored processed', and `user consent obtained storing'. From the above analysis, we can see that developers are aware that processing, storing and sharing are actions that require consent, even if the user has already opted-in to the general platform usage. Offline processing may have been proposed as a way to process data with reduced online exposure risks. 

\textbf{Scenario 3} (Privacy risks after deployment). The most commonly used word was `disable' with 50 counts. From the 2-gram phrases we can see an agreement on `auto disable' (20) and `disable unwanted' (7). From the 3-gram phrases we can combine answers like `tool auto disable' AND `disable unwanted features' and `tool disable unnecessary' to reveal that developers stated the need for an automated tool to detect such vulnerabilities and automatically disable them 31 times in total. We can also combine responses like `yes tool detect' AND `use tool auto' AND `definitely use tool' to identify that the participants stated that they would use such tool 15 times in total. 4-gram phrases are similar, but also reveal another important aspect. Some mentioned `detection alone isn't enough'(4). Developers clearly show the need for automated detection and mitigation tools. Repeating words like ` disable' and `auto disable' shows the need for preventing mechanisms and tools. We can observe strong interest on privacy-enforcing tools.    

\begin{figure}
    \centering
    \includegraphics[width=1\linewidth]{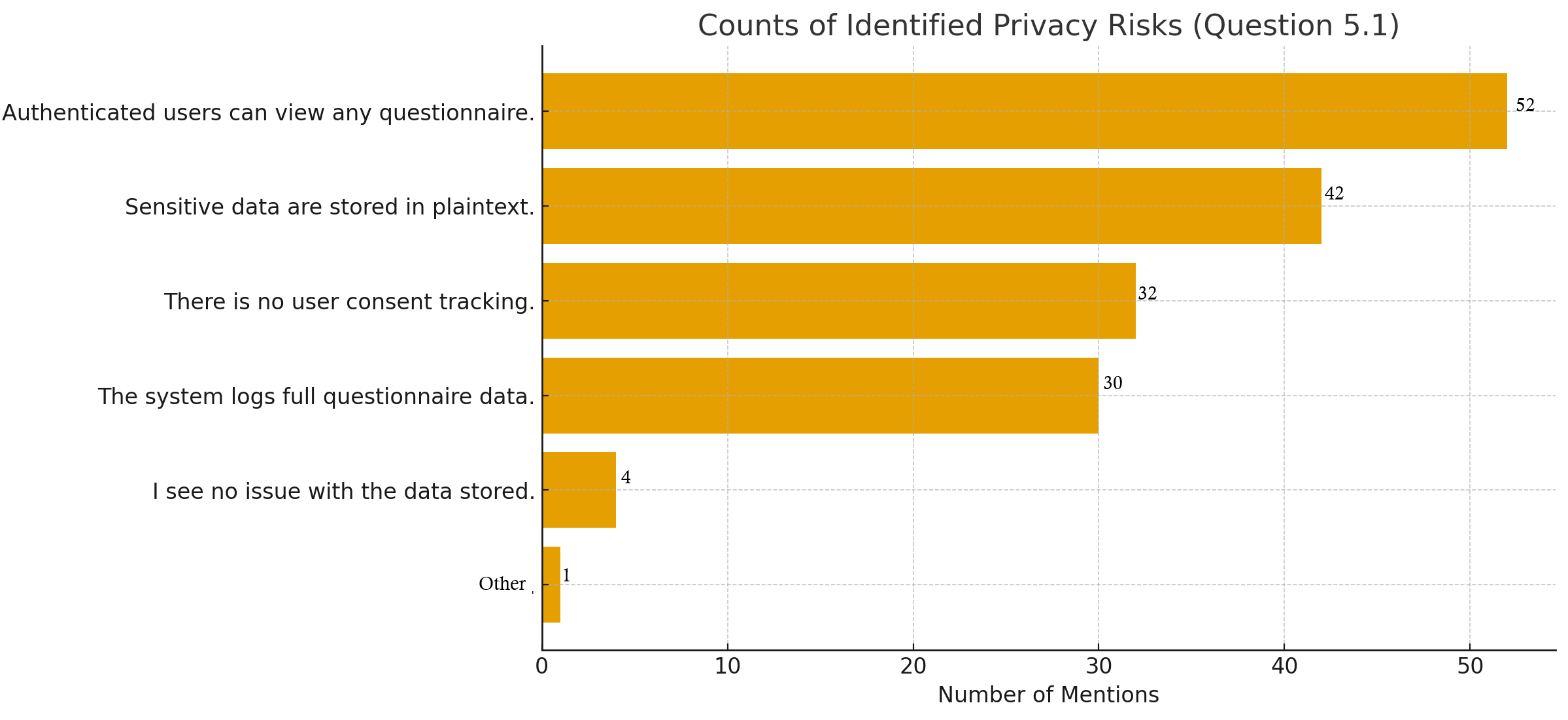}
    \caption{Counts of Identified Privacy Risks.}
    \label{fig:countsQ1s4}
\end{figure}

\textbf{Scenario 4} (Privacy as Code privacy leak detection). Question 1 presented a pool of privacy risks and developers were asked to select which risks they can identify in the code given in the scenario. Most voted response was that `\emph{authenticated users could view any questionnaire}' (Figure~\ref{fig:countsQ1s4}) which is the worst problem in that scenario along with the other vulnerabilities listed. Only four responses stated no issue with the code. Question 2 asked developers how they would incorporate privacy controls into the Agile workflow. Through the n-gram analysis, the two most popular words regarding solution proposals were: `planning' and `backlog'. The 2-grams showed `user stories', `acceptance criteria', `sprint planning', `ci cd' and `privacy checks' as the most frequent. 3-gram phrases showed `privacy user stories', `privacy acceptance criteria', `encrypt sensitive data' as proposed solutions and an important suggestion through 4-grams was `integrate privacy checks sprint'. Developers seem to recognize the importance of integrating privacy controls in Agile Software Development since they mentioned user stories, backlog items and acceptance criteria. Lastly, it seems that developers, by mentioning CI/CD, value automation processes.

\section{Implications and Threats to Validity}

\textbf{Main findings}. Overall, participants reported that there is a need for more privacy tools, especially automated. We found that participants who have experience with privacy (e.g. PETs) have a better understanding of privacy needs and thus, tend to report more needs for privacy tools. Participants from smaller companies did not find a lack on guidance available on PETs, and this might be attributed to the fact that there is less emphasis put on privacy due to limited resources and lack of dedicated roles for this. Concerning the scenarios, most participants provided valid answers and some patterns were found in top encountered terms: e.g. use of automated CI/CD processes, code analysis, pro-activeness, offline processing.

\textbf{Implications}. Our results show trends that can be followed by software engineers and researchers.\\
\emph{Practitioners}. A main take-away message for practitioners is that specific roles have more needs for privacy tools. They should bear in mind that if they are in such roles in the future, they will need to gain a deeper understanding of privacy compliance. According to our results, Product Managers have more awareness of documentation guidelines for privacy integration, so this might be expected more from their role.\\
\emph{Researchers}. Our work has shown that more tools are needed to support specifically professionals who have security/privacy engineering as their main task. Automated tools, reusable privacy design patterns and source code analyzers would be useful in this respect, while recent advances also show the need for more responsible Generative AI assistants in Privacy as Code~\cite{ferreyra2025good}.

\textbf{Threats to validity}. Our results are based on the analysis of 68 valid survey responses and may not generalize to a wider population of privacy practitioners, affecting \emph{external validity} of our work. We are relying on replies as self-reported by participants, and this might have affected \emph{construct validity}, in case the answers provided by the participants were not accurate. The same aspect could be affected by the lack of participants of all age groups. Finally, \emph{conclusions validity} is largely linked with the chosen statistical tests used to analyze our data.

\section{Conclusions}

In this work, we have performed a survey on current developers' privacy needs, with a focus on legislation compliance activities. We have analyzed 68 valid responses, and have examined the top general and technical needs, as well as how participants experience and demographics affect these needs. We have also reported the main replies participants gave to typical compliance-relevant scenarios. 
Our results demonstrate that practitioners are in need of more specific tools (e.g. design patterns for privacy, automated tools). As future work, we will work on the main needs indicated creating low-fidelity prototypes of the requested automated tools, putting mainly emphasis on offering automated tools that make compliance faster, relying on CodeLLMs for this process by e.g. providing them with more compliance-awareness~\cite{krishna2025codellm}.

\printbibliography

@article{eugdpr,
title={General Data Protection Regulation},
  author={{European Parliament and Council of the European Union}},
  year={2015},
  journal={Official Journal of the European Union}
}

@article{Leite2021,
    author = {Leite, Luís and dos Santos, Daniel Rodrigues and Almeida, Fernando},
    title = {The impact of general data protection regulation on software engineering practices},
    journal = {Information and Computer Security},
    volume = {30},
    number = {1},
    pages = {79-96},
    year = {2021},
    month = {08},
}

@article{Colesky2016,
  author    = {Colesky, Michael and Hoepman, Jaap-Henk and Hillen, Christiaan},
  title     = {A Critical Analysis of Privacy Design Strategies},
  journal   = {IEEE Security \& Privacy},
  volume    = {14},
  number    = {6},
  pages     = {33--41},
  year      = {2016},
  doi       = {10.1109/MSP.2016.111},
  url       = {https://ieeexplore.ieee.org/document/7762916}
}

@article{pallas2024privacy,
  title={Privacy engineering from principles to practice: A roadmap},
  author={Pallas, Frank and Koerner, Katharina and Barber{\'a}, Isabel and Hoepman, Jaap-Henk and Jensen, Meiko and Narla, Nandita Rao and Samarin, Nikita and Ulbricht, Max-R and Wagner, Isabel and Wuyts, Kim and others},
  journal={IEEE Security \& Privacy},
  volume={22},
  number={2},
  pages={86--92},
  year={2024},
  publisher={IEEE}
}

@inproceedings{prybylo2024evaluating,
  title={Evaluating privacy perceptions, experience, and behavior of software development teams},
  author={Prybylo, Maxwell and Haghighi, Sara and Peddinti, Sai Teja and Ghanavati, Sepideh},
  booktitle={Twentieth Symposium on Usable Privacy and Security (SOUPS 2024)},
  pages={101--120},
  year={2024}
}

@inproceedings{senarath2019Model,
author = {Senarath, Awanthika and Grobler, Marthie and Arachchilage, Nalin},
year = {2019},
month = {01},
pages = {},
title = {A Model for System Developers to Measure the Privacy Risk of Data},
doi = {10.24251/HICSS.2019.738}
}

@article{senarath2018PrExpect,
title = {Understanding user privacy expectations: A software developer’s perspective},
journal = {Telematics and Informatics},
volume = {35},
number = {7},
pages = {1845-1862},
year = {2018},
issn = {0736-5853},
doi = {https://doi.org/10.1016/j.tele.2018.05.012},
url = {https://www.sciencedirect.com/science/article/pii/S073658531830296X},
author = {Awanthika R. Senarath and Nalin Asanka Gamagedara Arachchilage},
}

@inproceedings{franke2024exploratory,
author = {Franke, Lucas and Liang, Huayu and Farzanehpour, Sahar and Brantly, Aaron and Davis, James C. and Brown, Chris},
title = {An Exploratory Mixed-methods Study on General Data Protection Regulation (GDPR) Compliance in Open-Source Software},
year = {2024},
isbn = {9798400710476},
publisher = {Association for Computing Machinery},
address = {New York, NY, USA},
url = {https://doi.org/10.1145/3674805.3686692},
doi = {10.1145/3674805.3686692},
pages = {325–336},
numpages = {12},
location = {Barcelona, Spain},
series = {ESEM '24}
}

@INPROCEEDINGS{ferreyra2023SCPs,
  author={Ferreyra, Nicolás E. Díaz and Imine, Abdessamad and Vidoni, Melina and Scandariato, Riccardo},
  booktitle={2023 IEEE/ACM 16th International Conference on Cooperative and Human Aspects of Software Engineering (CHASE)}, 
  title={Developers Need Protection, Too: Perspectives and Research Challenges for Privacy in Social Coding Platforms}, 
  year={2023},
  volume={},
  number={},
  pages={105-110},
  doi={10.1109/CHASE58964.2023.00019}}

@article{trujillo19,
author = {Trujillo, Miguel Ehecatl and García-Mireles, Gabriel and Matla Cruz, Erick Orlando and Piattini, Mario},
year = {2019},
month = {04},
pages = {},
title = {A Systematic Mapping Study on Privacy by Design in Software Engineering},
volume = {22},
journal = {CLEI Electronic Journal},
doi = {10.19153/cleiej.22.1.4}
}

@inproceedings{mao2024privacycat,
  author    = {K. Mao and C. T. Åhs and S. Cela and D. Distefano and N. Gardner and R. Grigore and P. Gustafsson and Á. Hajdu and T. Kapus and M. Marescotti and G. C. Sampaio and T. Suzanne},
  title     = {PrivacyCAT: Privacy-Aware Code Analysis at Scale},
  booktitle = {46th International Conference on Software Engineering: Software Engineering in Practice (ICSE-SEIP ’24)},
  year      = {2024},
  address   = {Lisbon, Portugal},
  publisher = {ACM}
}

@inproceedings{Tahaei20,
title = "Understanding Privacy-Related Questions on Stack Overflow",
author = "Mohammad Tahaei and Kami Vaniea and Naomi Saphra",
year = "2020",
month = apr,
day = "21",
doi = "10.1145/3313831.3376768",
language = "English",
isbn = "9781450367080",
booktitle = "CHI '20: Proceedings of the 2020 CHI Conference on Human Factors in Computing Systems",
publisher = "ACM",
note = "ACM CHI Conference on Human Factors in Computing Systems, CHI 2020 ; Conference date: 25-04-2020 Through 30-04-2020",
url = "https://chi2020.acm.org/",
}

@article{alhazmi2021gdpr,
  author    = {A. Alhazmi and N. A. G. Arachchilage},
  title     = {I’m All Ears! Listening to Software Developers on Putting GDPR Principles into Software Development Practice},
  journal   = {Personal and Ubiquitous Computing},
  publisher = {Springer},
  year      = {2021},
  pages     = {879--892}
}

@article{Iwaya_2023,
   title={Privacy Engineering in the Wild: Understanding the Practitioners’ Mindset, Organizational Aspects, and Current Practices},
   volume={49},
   ISSN={2326-3881},
   url={http://dx.doi.org/10.1109/TSE.2023.3290237},
   DOI={10.1109/tse.2023.3290237},
   number={9},
   journal={IEEE Transactions on Software Engineering},
   publisher={Institute of Electrical and Electronics Engineers (IEEE)},
   author={Iwaya, Leonardo Horn and Babar, Muhammad Ali and Rashid, Awais},
   year={2023},
   month=sep, pages={4324–4348} }

@inproceedings{Tahaei23,
author = {Tahaei, Mohammad and Abu-Salma, Ruba and Rashid, Awais},
title = {Stuck in the Permissions With You: Developer \& End-User Perspectives on App Permissions \& Their Privacy Ramifications},
year = {2023},
isbn = {9781450394215},
publisher = {Association for Computing Machinery},
address = {New York, NY, USA},
url = {https://doi.org/10.1145/3544548.3581060},
doi = {10.1145/3544548.3581060},

booktitle = {Proceedings of the 2023 CHI Conference on Human Factors in Computing Systems},
articleno = {168},
numpages = {24},
location = {Hamburg, Germany},
series = {CHI '23}
}

@article{ Hadar18,
author = {Hadar, Irit and Hasson, Tomer and Ayalon, Oshrat and Toch, Eran and Birnhack, Michael and Sherman, Sofia and Balissa, Arod},
year = {2018},
month = {02},
pages = {},
title = {Privacy by designers: software developers’ privacy mindset},
volume = {23},
journal = {Empirical Software Engineering},
doi = {10.1007/s10664-017-9517-1}
}

@article{ Deng11,
author = {Deng, Mina and Wuyts, Kim and Scandariato, Riccardo and Preneel, Bart and Joosen, Wouter},
year = {2011},
month = {03},
pages = {3-32},
title = {A privacy threat analysis framework: Supporting the elicitation and fulfillment of privacy requirements},
volume = {16},
journal = {Requir. Eng.},
doi = {10.1007/s00766-010-0115-7}
}

@inproceedings{Senarath18Investigation,
author = {Senarath, Awanthika and Arachchilage, Nalin A. G.},
title = {Why developers cannot embed privacy into software systems? An empirical investigation},
year = {2018},
isbn = {9781450364034},
publisher = {Association for Computing Machinery},
address = {New York, NY, USA},
url = {https://doi.org/10.1145/3210459.3210484},
doi = {10.1145/3210459.3210484},
booktitle = {Proceedings of the 22nd International Conference on Evaluation and Assessment in Software Engineering 2018},
pages = {211–216},
numpages = {6},
location = {Christchurch, New Zealand},
series = {EASE '18}
}

@article{Hasani23,
author = {Tahereh Hasani and Davar Rezania and Nadège Levallet and Norman O’Reilly and Mohammad Mohammadi},
title ={Privacy enhancing technology adoption and its impact on SMEs’ performance},

journal = {International Journal of Engineering Business Management},
volume = {15},
number = {},
pages = {18479790231172874},
year = {2023},
doi = {10.1177/18479790231172874},

URL = { 
    
        https://doi.org/10.1177/18479790231172874
    
    
}}

@inproceedings{Tahaei2022understanding,
author = {Tahaei, Mohammad and Li, Tianshi and Vaniea, Kami},
year = {2022},
month = {07},
pages = {},
title = {Understanding Privacy-Related Advice on Stack Overflow}
}

@article{GUBER2024107351,
title = {Privacy-Compliant Software Reuse in Early Development Phases: A Systematic Literature Review},
journal = {Information and Software Technology},
volume = {167},
pages = {107351},
year = {2024},
issn = {0950-5849},
doi = {https://doi.org/10.1016/j.infsof.2023.107351},
url = {https://www.sciencedirect.com/science/article/pii/S0950584923002069},
author = {Jenny Guber and Iris Reinhartz-Berger}
}

@article{Kühtreiber22,
author = {Kühtreiber, Patrick and Pak, Viktoriya and Reinhardt, Delphine},
year = {2022},
month = {09},
pages = {101656},
title = {A survey on solutions to support developers in privacy-preserving IoT development},
volume = {85},
journal = {Pervasive and Mobile Computing},
doi = {10.1016/j.pmcj.2022.101656}
}

@article{ Sangaroonsilp25,
author = {Sangaroonsilp, Pattaraporn and Dam, Hoa Khanh},
title = {A Study on the Prevalence of Privacy in Software Engineering},
year = {2025},
issue_date = {November 2025},
publisher = {Association for Computing Machinery},
address = {New York, NY, USA},
volume = {57},
number = {11},
issn = {0360-0300},
url = {https://doi.org/10.1145/3734216},
doi = {10.1145/3734216},
journal = {ACM Comput. Surv.},
month = jun,
articleno = {288},
numpages = {34}
}

@article{Altschuld22,
author = {James W. Altschuld and Hsin-Ling (Sonya) Hung and Yi-Fang Lee},
title ={What Is and What Should Be Needs Assessment Scales: Factors Affecting the Trustworthiness of Results},

journal = {American Journal of Evaluation},
volume = {43},
number = {4},
pages = {607-619},
year = {2022},
doi = {10.1177/10982140211017663},

URL = { 
    
        https://doi.org/10.1177/10982140211017663
    
}
}

@inproceedings{krishna2025codellm,
  title={Codellm-devkit: A framework for contextualizing code llms with program analysis insights},
  author={Krishna, Rahul and Pan, Rangeet and Sinha, Saurabh and Tamilselvam, Srikanth and Pavuluri, Raju and Vukovic, Maja},
  booktitle={Proceedings of the 33rd ACM International Conference on the Foundations of Software Engineering},
  pages={308--318},
  year={2025}
}

@inproceedings{ferreyra2025good,
  title={The Good, the Bad, and the (Un) Usable: a Rapid Literature Review on Privacy as Code},
  author={Ferreyra, Nicol{\'a}s E D{\'\i}az and Khelifi, Sirine and Arachchilage, Nalin and Scandariato, Riccardo},
  booktitle={2025 IEEE/ACM 18th International Conference on Cooperative and Human Aspects of Software Engineering (CHASE)},
  pages={173--178},
  year={2025},
  organization={IEEE}
}

@inproceedings{vanezi2020dialogop,
  title={Di{\'a}logop-A language and a graphical tool for formally defining GDPR purposes},
  author={Vanezi, Evangelia and Kapitsaki, Georgia M and Kouzapas, Dimitrios and Philippou, Anna and Papadopoulos, George A},
  booktitle={International Conference on Research Challenges in Information Science},
  pages={569--575},
  year={2020},
  organization={Springer}
}
\end{document}